\documentclass[12pt,preprint]{aastex}

\slugcomment{submitted to ApJ.}

\shorttitle{Helium settling and rotation-induced mixing : numerical approach}
\shortauthors{Sylvie Th\'eado and Sylvie Vauclair}

\begin{document}

\title{On the coupling between helium settling and rotation-induced
mixing in stellar radiative zones : II- Numerical approach}

\author{Sylvie Th\'eado and Sylvie Vauclair}
\affil{Laboratoire d'Astrophysique, Observatoire Midi-Pyr\'en\'ees, 14
avenue Edouard Belin, 31400 Toulouse, France}

\begin{abstract}
In the first paper of this series, we discussed analytically, in an
approximate way, the mixing processes which occur in slowly rotating
stars when the feed-back effects due to the diffusion-induced
$\mu$-gradients are introduced in the computations. We found that the
classical scheme of meridional circulation was dramatically modified,
with important possible consequences. Here we present a complete 2D
numerical simulation of these processes in a static stellar model. The
helium microscopic diffusion and the meridional advection are computed
simultaneously, so that we can follow the evolution of the abundances
and mixing processes in the radiative zone. Most of the effects
discussed in paper I are confirmed. Some consequences will be discussed
in a forthcoming paper (Th\'eado and Vauclair 2002, paper III).
\end{abstract}


\section{Introduction}
This paper is the second of a series concerning the importance of
diffusion-induced $\mu$-gradients in the computations of
rotation-induced mixing in stellar radiative zones. In the first paper
of this series (Vauclair and Th\'eado 2002, paper I), we discussed in
details the orders of magnitude of the various terms which appear in the
computation of the meridional circulation velocity. We showed that the
term related to the transport of angular momentum may be important in
some cases and particularly at the beginning of the main-sequence for F
and G type stars. However, during most of their lifetime, their
hydrodynamical history is dominated by the competition between the
classical terms (the Eddington-Sweet meridional circulation) and the
terms related to the feed-back effect induced by $\mu$-gradients. 

The situation which appears when these terms become of the same order of
magnitude and nearly cancel out is difficult to handle numerically. In
paper I we gave an approximate analytical approach to better understand,
in a physical way, the hydrodynamical processes which may happen when
this situation takes place. We found that a quasi-stationary stage can
settle down, in which case the circulation pattern is strongly modified.
The vertical and horizontal $\mu$-gradients ajust themselves to remain
nearly constant in the concerned zone. Such a situation must be taken
into account in the computations of abundance variations in stellar
outer layers.

In the past, various authors performed 2D numerical simulation of meridional circulation (Tassoul and Tassoul \cite{Tassoul82}, Mestel and Moss \cite{Mestel86}, Charbonneau \cite{Charbonneau92}). In the computations of Tassoul and Tassoul \cite{Tassoul82} and Mestel and Moss \cite{Mestel86} the importance of the feed-back effect due to nuclearly-induced $\mu$-gradients was recognized. In particular Mestel and Moss \cite{Mestel86} showed that the $\mu$-currents
could slowly stabilize the circulation near the core. They called this process ``creeping paralysis''. However in these computations the feed-back effect due to the diffusion-induced $\mu$-gradients was not included and the horizontal and vertical composition variations induced by the diffusion/circulation coupling were not accurately studied.

In this paper, we present the results of a 2D numerical simulation of
meridional circulation in the presence of helium settling, including the
feed-back effect due to $\mu$-gradients.
In order to calculate and
visualize in a realistic way the physical processes which take place
below the convective zone, we have computed simultaneously the helium
abundance variations and the meridional circulation velocity. No free
parameter is introduced in these computations, although some simplifying
assumptions are made : the rotation velocity is assumed constant during
the time of the simulation, and the modifications of the stellar
structure induced by diffusion and evolution are neglected (static
model). In the simulation, the meridional circulation is treated as an
advection process. However, as will be discussed in section 2.2, the
discretization leads to some numerical diffusion, which we evaluate and
find similar to an anisotropic mixing, larger horizontally than
vertically, in a realistic way.

In section 2 we give a complete description of the numerical method. The
simulation is first tested on the well-known classical circulation
(section 3) and the results obtained including the effects of
$\mu$-gradients are given in section 4. A general discussion of these
results is proposed in section 5.

\section{Numerical analysis}
We describe below the details of our computational method. 
Throughout this paper, we use the spherical coordinate system but we
assume axial symetry so that the studied processes are reduced to a 2D
simulation.

\subsection{Discretization}
\begin{figure}[h]
\epsscale{0.3}
\plotone{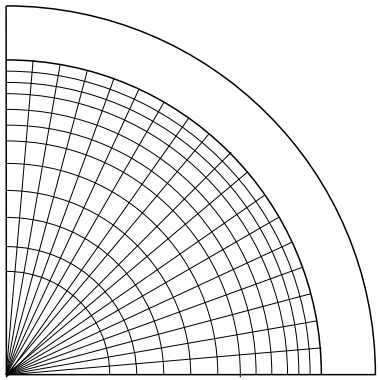}
\caption{Discretization of a meridional plane quadrant (schematic representation). In the computations the model is
divided in about 800 shells in the vertical direction and
 90 angular sectors in the horizontal direction.
The meshpoint is tighter in the vertical than in the horizontal
direction.}
\label{maille}	
\end{figure}

As an initial model for our simulation, we used a standard homogeneous
stellar model (here we chose a 0.75M$_{\odot}$ halo star, which has no
importance for the general process we intent to describe). To obtain the
2D meshpoint of the simulation, each shell of the 1D initial model is
first discretized in the latitudinal direction into angular sectors of
one arcdeg (figure \ref{maille}).

In order to treat in a more accurate way the advection by meridional
circulation and to reduce the numerical diffusion, a secondary meshpoint
will be defined : each cell of the main meshpoint will be divided into
smaller cells. We will come back on this secondary meshpoint in section
\ref{na:meridcirc}.

\subsection{Computational method}
This simulation aims to understand the coupling between microscopic
diffusion and meridional circulation. These two physical processes have
quite different natures and behaviors. Microscopic diffusion is a
particle process which leads to a selective transport of chemical
elements. On the other hand, meridional circulation is a hydrodynamical
process which leads to a global transport of matter. 
We have included in the code two different routines which compute
each of the two transport processes in an adequate way :

\begin{itemize}
\item  a particle routine which first computes the microscopic diffusion
of helium with an eulerian scheme,
\item  a hydrodynamical routine which includes a lagrangian description
of the transport process to compute the global advection by meridional
circulation.
\end{itemize}

At each timestep, we first compute the diffusion-induced composition
variations by using the particle routine. Then, taking into account the
new chemical composition, we determine the circulation-induced abundance
variations by computing the advection of matter with the hydrodynamical
routine.

The timesteps always satisfy the flow condition. Using smaller timesteps does not alter the results. We have also checked that reversing the order of the diffusion and circulation computations does not alter the results in a significant way.

\subsubsection{The treatment of microscopic diffusion}
The diffusion-induced composition variations are computed by solving the
helium mass conservation equation:
\begin{equation}
 \frac{\partial{\rho Y}}{\partial t} + div(\rho \mathbf{V} Y) =0
\end{equation}
where Y represents the helium mass fraction and $\mathbf{V}$ the helium
diffusion velocity. In the region below the convective zone, the helium
production rate by nuclear reactions is negligible. 

The diffusion velocity is computed in the test-atom approximation with respect to hydrogen :
\begin{equation}
\mathbf{V}=D_{He}\left\{ - \mathbf{\nabla} \ln c_{He} + k_p \mathbf{\nabla} \ln p +k_T \mathbf{\nabla}
\ln T +\frac{m\mathbf{F}}{kT}\right\}
\end{equation}
$D_{He}$ is the helium diffusion coefficient in a hydrogen gas, which we
compute with the Paquette et al. (1986) formalism, and the four terms in
the bracket represent respectively the concentration gradient, the
gravitational settling, the thermal and the radiative diffusion.

The test-atom approximation is rather crude for helium, whose abundance
is about 10\% that of hydrogen. However, for the purpose of the
present paper, treating helium diffusion with a more precise
approximation would have lead to useless heavier computations in the 2D
simulation with a difference in the results of 10 to 15 \% (Montmerle and Michaud \cite{Montmerle76}).

In solar type stars, the radiative acceleration is negligible. In the vertical direction, the diffusion velocity is then due to the concentration gradient, the
gravitational settling and the thermal diffusion. In the horizontal direction the velocity depends only on the concentration gradient. 

The horizontal and the vertical component of $\nabla \ln c_{He}$ are respectively equal to : $\displaystyle \frac{1}{r} \frac{1}{c}\frac{\partial c}{\partial \theta}$ and $\displaystyle \frac{1}{c} \frac{\partial c}{\partial r}$.
As a first approximation, the helium concentration is a linear function of the inverse of the molecular weight so $\displaystyle \frac{|dc|}{c}$ can be approximated by $\displaystyle \frac{|d\mu|}{\mu}$. 

In the horizontal direction, we will see in section 4 that in the considered stars $\displaystyle \frac{d\mu}{\mu}$ is typically of the order of 10$^{-6}$. With a radius of about 5$\cdot$10$^{10}$ cm, it leads to the following approximation :
$$\vert\frac{1}{r} \frac{1}{\mu} \frac{\partial \mu}{\partial \theta}\vert\simeq 2 \cdot 10^{-17} \rm{cm}^{-1}$$
This leads to a horizontal diffusion velocity of nearly 8$\cdot$10$^{-17}$ cm.s$^{-1}$ while the horizontal meridional velocity is typically of the order of 10$^{-8}$ to 10$^{-9}$ cm.s$^{-1}$ : the horizontal diffusion can therefore be neglected in the computations. 

In the vertical direction $\displaystyle \vert \frac{1}{c} \frac{\partial c}{\partial r}\vert$ can be approximated by $\displaystyle\vert \frac{1}{\mu} \frac{\partial \mu}{\partial r}\vert$. We will see in section 3.1 that just below the convective zone, where the $\mu$-gradients are the steepest :  $$ \vert\frac{1}{\mu} \frac{d\mu}{dr}\vert \simeq 9 \cdot 10^{-12} \rm{cm}^{-1} $$
It is interesting to compare the concentration gradient contribution with the others terms and in particular the gravitational settling term. In the case of a totally ionised gas, this term can be written :
$$k_p \nabla \ln p=\frac{5}{2} \frac{m_p}{kT}\frac{ G m_r}{r^2}$$
In a 0.75M$_{\odot}$ (i.e. a radius of nearly 5$\cdot$10$^{10}$ cm), this term is approximately 1.7$\cdot$10$^{-9}$ cm.s$^{-1}$. Typically the vertical concentration gradient term is at least 2 order of magnitude smaller than the other terms.

Two boundary conditions are needed. The upper one is obtained by
computing the dilution in the convective zone.	
\begin{equation}
\int_{cz} \frac{\partial \rho X}{\partial t} dv + \int_{cz} \rho X
\mathbf{V.dS} =0 
\end{equation}
which can be written in quasi-stationary regime :
\begin{equation}
M_c \frac{\partial X}{\partial t}= 2 \pi \int_0^{\pi} \rho X V_r r^2
\sin \theta d\theta 
\end{equation}
where $M_c$ is the stellar convective mass.

The nuclear reaction rates are neglected and we assume that the
diffusion time scale is much greater than the stellar life : in these
conditions, the lower boundary condition is chosen so that $Y=$constant
below $r \simeq$ 0.4 R$_{\odot}$. 

	\subsubsection{The treatment of meridional circulation}
\label{na:meridcirc}
The vertical component of the meridional circulation is traditionally
expanded in spherical functions. When the rotation rate depends only on
depth, the radial velocity involves only the second Legendre polynomial
: 
\begin{equation}
u_{r} =U_{r} \ P_{2} \ (\cos \theta)
\end{equation}
The velocity amplitude, $U_{r}$, is derived from the heat transfert
equation. It can be written (Vauclair and Th\'eado 2002, paper I):
\begin{equation}
U_{r}=\frac{P}{\rho g T C_p (\nabla_{ad}-\nabla+\nabla_{\mu})}
\frac{L}{M_*} (E_{\Omega}+E_{\mu}+E_{\zeta}+E_h)
\label{Ur}
\end{equation}
$\nabla_{ad}$ and $\nabla$ represent the usual adiabatic and real ratios
$\displaystyle \left( {d \ln T\over d \ln P }\right)$, $\nabla_{\mu}$
the mean molecular weight contribution $\displaystyle \left( {d \ln \mu
\over d \ln P}
\right)$. The four terms in the bracket represent :
\begin{itemize}
\item the classical Eddington-Sweet term $E_{\Omega}$ :
\begin{equation}
E_{\Omega} = 2 \left[ 1-\frac{\bar{\Omega}^2}{2 \pi G \bar{\rho}}
\right] \frac{\tilde g}{\bar{g}} 
  - \frac{\rho_m}{\bar {\rho}} \left\{ \frac{r}{3}\frac{d}{dr} \left[
H_T \frac{d\zeta}{dr} -\chi_T\zeta \right]
-\frac{2 H_T}{r} 
  \zeta +\frac{2}{3}\zeta \right\} 
 - \frac{\bar{\Omega}^2}{2 \pi G \bar{\rho}} \zeta 
\end{equation}
Note that the last term in this equation was not written in the previous
papers : it was neglected by mistake in Maeder and Zahn 1998. While it
is indeed negligible in the deep stellar regions where the so-called
``Gratton-$\ddot{O}$pik term" is less than unity, it can become important in the
outer layers. 
\item the $\mu$-gradient term $E_{\mu}$ :
\begin{equation}
E_{\mu}=\frac{\rho_m}{\bar{\rho}}\left\{ \frac{r}{3} \frac{d}{dr} \left[
H_T
\frac{d\Lambda}{dr}  - \left(
\chi_{\mu}+\chi_T+1 \right)
\Lambda \right] - 2\frac{H_T}{r}\Lambda \right\} 
\label{emu} 
\end{equation}
\item the term related to the differential rotation $E_{\zeta}$ :
\begin{equation}
E_{\zeta}=\frac{M_*}{L}\bar{T} C_p \frac{\partial \zeta}{\partial t}
\end{equation}
\item the term related to horizontal turbulence $E_h$ : 
\begin{equation}
E_h=\frac{\rho_m}{\bar{\rho}} \frac{2 H_T}{r} \frac{D_h}{K}\zeta
\end{equation}
\end{itemize}

In these equations, $\zeta$ represents the density fluctuations along a level surface $\displaystyle \frac{\tilde{\rho}}{\bar{\rho}}$ ; $\displaystyle \Lambda$  refers to the horizontal $\displaystyle \mu$-fluctuations $\displaystyle {\tilde{ \mu}\over \overline \mu }$ ; $\rho_m$ is the mean density inside the sphere of radius r while $\bar{\rho}$ represents
the density average on the level surface (as well as $\bar{T}$ for the
temperature and $\bar{\Omega}$ for the angular rotation velocity) ;
$C_p$ is the specific heat ; $H_T$ the temperature scale height and
$D_h$ the horizontal turbulent diffusion coefficient.
$\chi_{\mu}$ and $\chi_{T}$ are given by :
$$
\chi_{\mu } =
\left(
{\partial \ln \chi \over \partial \ln \mu  }\right)_{P,T}
\quad  ; \quad
\chi_{T} =
\left( {\partial \ln \chi \over \partial \ln T }\right)_{P, \mu }
$$

In paper I, we showed that the term $E_h$ due to horizontal turbulence
is always negligible compared to the other terms. We also showed that
the term $E_\zeta$, which is related to the transport of angular
momentum, may be important at the beginning of the stellar lifetime. We
neglect it here however, as our purpose is to study the behavior of the
circulation when the two other terms $E_\Omega$ and $E_\mu$ are
preponderant.

The horizontal velocity component of the meridional circulation is then
deduced from the equation of matter conservation:
\begin{equation}
div (u_r \rho r^2)=- u_{\theta} \rho r 
\end{equation}
which gives :
\begin{equation}
u_{\theta} =
- {1 \over  2 \rho r} 
{d \over dr}\
(\rho r^{2} U_{r})
\sin \theta \cos \theta
\label{ut}
\end{equation}

\subsubsection{The numerical method}
\begin{figure}[h]
\epsscale{0.8}
\plotone{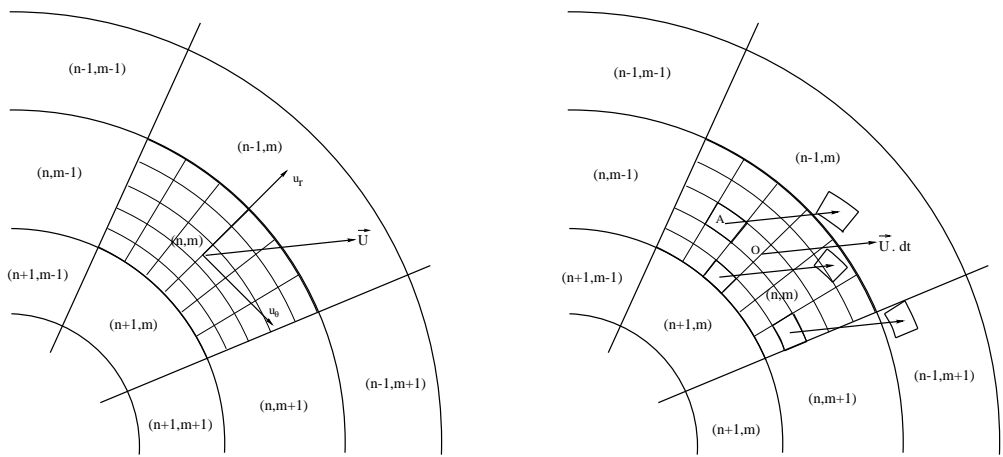}
\caption{Schematic representation of the meridional circulation
advection : secondary meshpoint and cell advection.}
\label{maillefin}	
\end{figure}

We use the finite difference technique to discretize the equations. The
timestep of the simulation is computed in order to satisfy the flow
condition. At each timestep, the microscopic diffusion is first computed
by determining the helium diffusion velocity in each shell and solving
the helium mass conservation equation for each cell of the meshpoint.
Once the diffusion-induced abundance variations are computed in the
whole star, the laminar meridional circulation is then treated as an
advection process. The velocity components are first determined at the
center of each cell. Then, in order to treat in a more accurate way the
advection by circulation and to reduce the numerical diffusion, a
secondary meshpoint is introduced in the computations. Each cell of the
main meshpoint (described in figure \ref{maille}) is discretized into 10$^4$ 
smaller cells. All the secondary cells belonging to the same main cell are
supposed to be advected with the same velocity : the one computed at the
center of the main cell. This assumption introduced an error lower than 5\% on the velocity components. The matter contained in each cell of the
secondary meshpoint is supposed to be concentrated at its center. The
motion of each secondary cell is computed and the contained matter is
distributed into the main cell in an adequate way (see figure
\ref{maillefin}). The main cells are then homogeneized to obtain the new
chemical composition in the star. 

\subsection{Numerical diffusion}
In spite of the secondary meshpoint introduced in the computations of
meridional circulation, the numerical scheme suffers some numerical
diffusion. This is basically due to the fact that, at each time step,
matter is homogeneized inside each cell of the main meshpoint. 

The numerical diffusion cannot be totally suppressed, however it can be reduced by increasing the number of cells. In practice increasing the number of cells leads rapidly to extremely large computations time which are unacceptable. The 800$\times$90 chosen meshpoint leads to both reasonable computation times and reasonable numerical diffusion. 

It is possible to evaluate this diffusion by defining an anisotropic
numerical diffusion coefficient which we call $D_h^{num}$ in the
horizontal direction and $D_v^{num}$ in the vertical direction. In both
case it is defined as 
$D^{num} = {dr}^2 / dt$ where dr is the length scale of the cell and dt
the time step of the computation.

In the horizontal direction $dr \simeq r/100$ while in the vertical one
$dr \simeq r/1000$ . The time steps of the simulation lie between $10^6$
and $10^7$ yr. With a radius of about $5 \cdot 10^{10}$ cm, we find the
orders of magnitude : $D_h^{num} \simeq 10^3$ to $10^4$ cm$^2$.s$^{-1}$
and $D_v^{num} \simeq 10$ to $100$ cm$^2$.s$^{-1}$.

It is interesting to compare the value of the numerical horizontal
diffusion coefficient with the prescription given by Zahn (1992) and
Maeder and Zahn (1998): 
$D_h = C_h r \cdot U_r$ . With $U_r$ of order $10^{-7}$ cm.s$^{-1}$, we
find $C_h \simeq 0.2$ to $2$, which is quite reasonable. Meanwhile the
numerical vertical diffusion coefficient is $100$ times smaller. 

We thus find that our numerical scheme leads to some small artificial
mixing, which simulates in a realistic way the kind of turbulence which
may occur physically in such a situation.

\section{Test of the numerical scheme}
To test the numerical scheme, we first check that the simulation
reproduces correctly the processes we already know, like microscopic
diffusion alone and classical meridional circulation, without the
$\mu$-terms. Then we introduce the feed-back effect due to the
$\mu$-gradients to try to understand more precisely what happens in
those circumstances.

\subsection{The classical meridional circulation}
We first introduce in the simulation the diffusion of helium and the
advection by the $\Omega$-currents only : the effects of $\mu$-gradients
are not taken into account, which means that the $E_{\mu}$ term is not
introduced in the computations of the meridional velocity.
\begin{figure}[h]
\epsscale{0.4}
\plotone{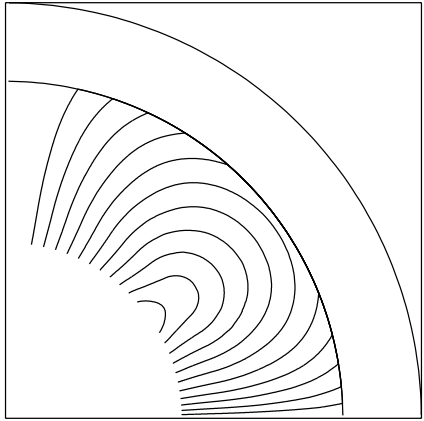}
\caption{Meridional circulation streamlines : here, only the
$\Omega$-currents are taken into account in the meridional velocity
computations. The rotation velocity is chosen constant during the whole
simulation and no structural evolutions are introduced in the
computations, therefore the $\Omega$-currents streamlines remain the
same at each timestep.}
\label{ligneEom}
\end{figure}

\begin{figure}[h]
\epsscale{0.8}
\plottwo{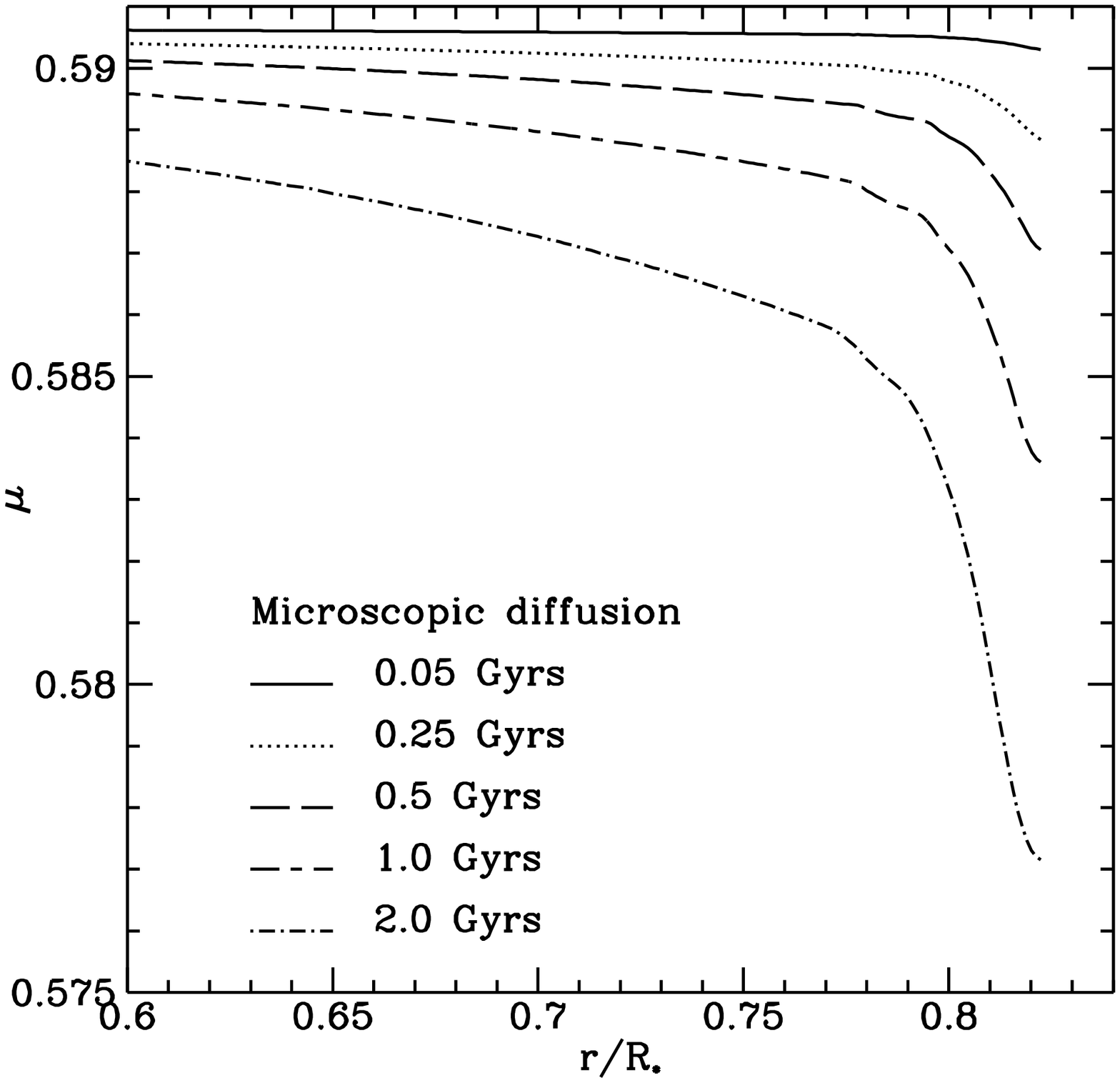}{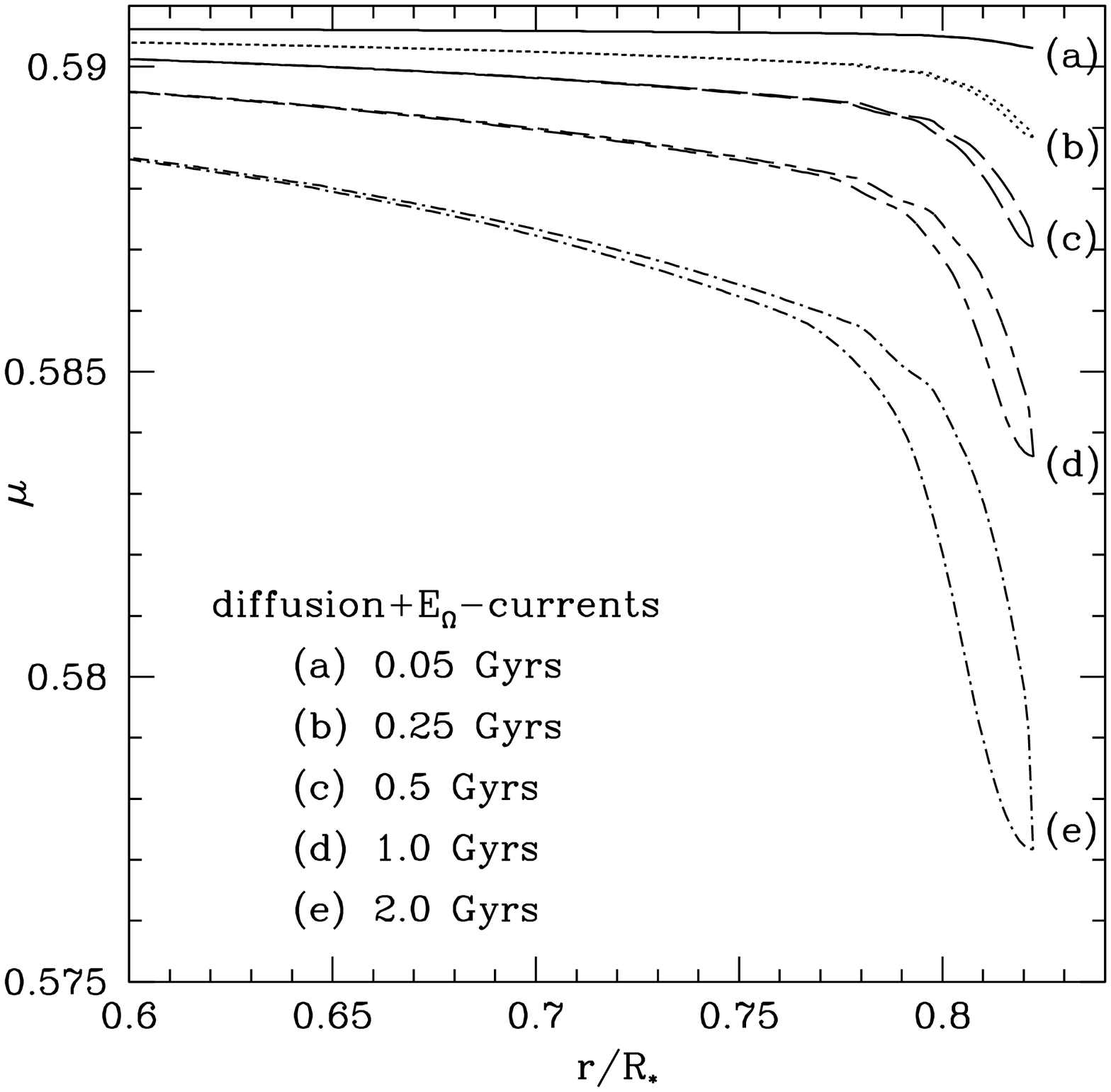}
\caption{Molecular weight versus radius in each cell below the
convective zone.
The left figure shows the results obtained when helium microscopic
diffusion is the only transport process taken into account. Under the
effects of microscopic diffusion vertical $\mu$-gradients build below
the convective zone. The horizontal microscopic diffusion is negligeable
so all the cells of a same shell have the same chemical composition and
the same $\mu$ value.  
The right figure displays the molecular weight 
when the diffusion of helium and the classical meridional
circulation (i.e. the $\Omega$-currents) are introduced in the
simulation. For each age two extreme curves are given. They represent at a fixed radius the spread of the $\mu$-values in the horizontal direction.
The $\Omega$-currents bring $\mu$-enriched matter up to the
rotation axis and down to the equator. This turns the vertical
diffusion-induced $\mu$-gradients into horizontal ones which explains
the spread in the value of $\mu$ at a fixed radius.}
\label{murdiff}
\end{figure}

Here we present the results obtained for a constant rotation velocity of
5 km.s$^{-1}$. The circulation streamlines are represented on figure
\ref{ligneEom}. As expected, the circulation leads to ascending flows
near the rotation axis and to descending flows near the equator. The
flow brings $\mu$-enriched matter up in the polar axis and
$\mu$-depleted matter down in the equatorial regions . Figure
\ref{murdiff} displays the molecular weight profiles at different steps
of the simulation. For comparison it also shows the results obtained
when helium diffusion is the only transport process introduced in the
computations (i.e. without meridional circulation). Microscopic
diffusion alone leads to an important vertical $\mu$-gradient below the
convective zone. The spread of the molecular weight values obtained in
the presence of meridional circulation shows how $\Omega$-currents turn
the diffusion-induced vertical $\mu$-gradients into horizontal ones.

\begin{figure}[h]
\epsscale{0.7}
\plotone{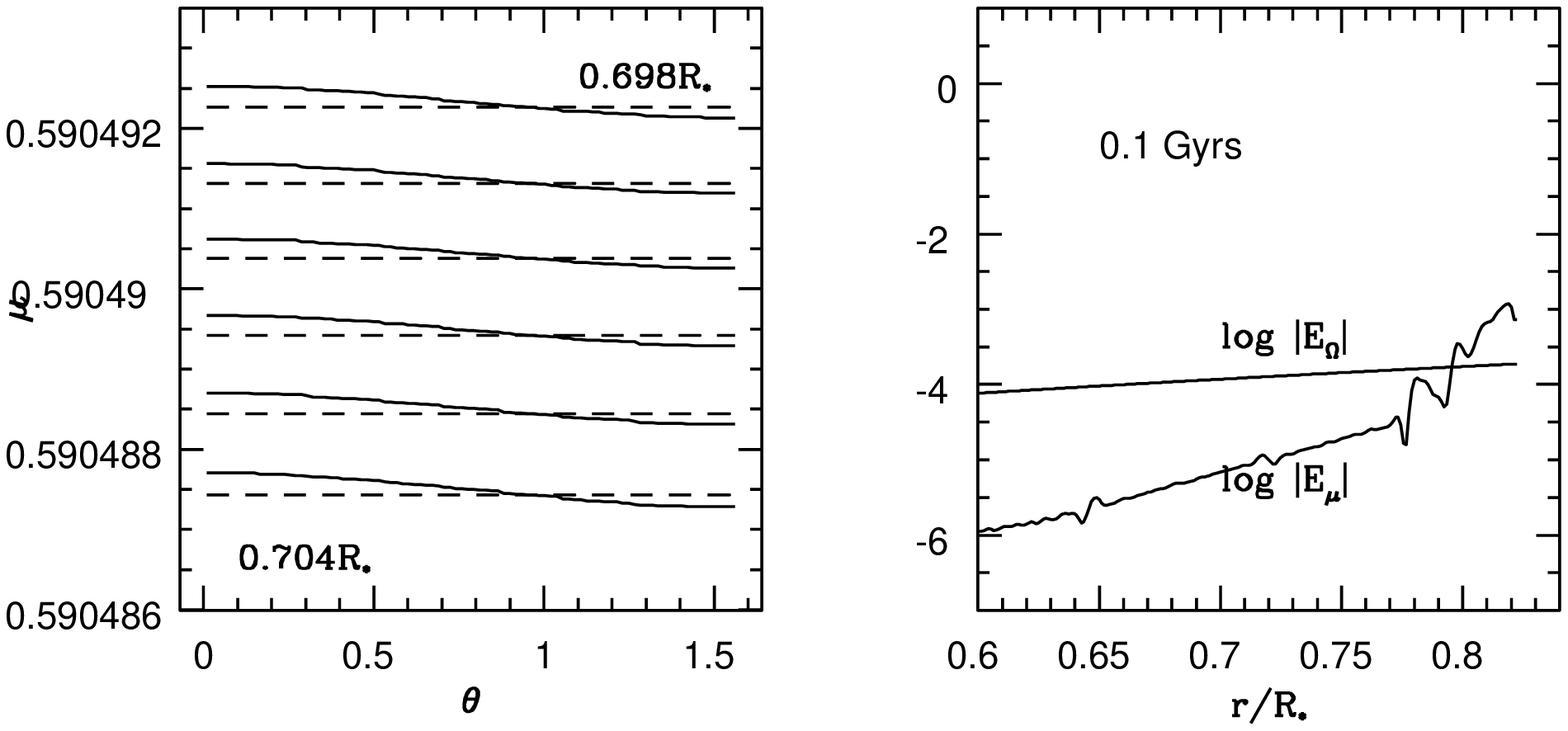}
\plotone{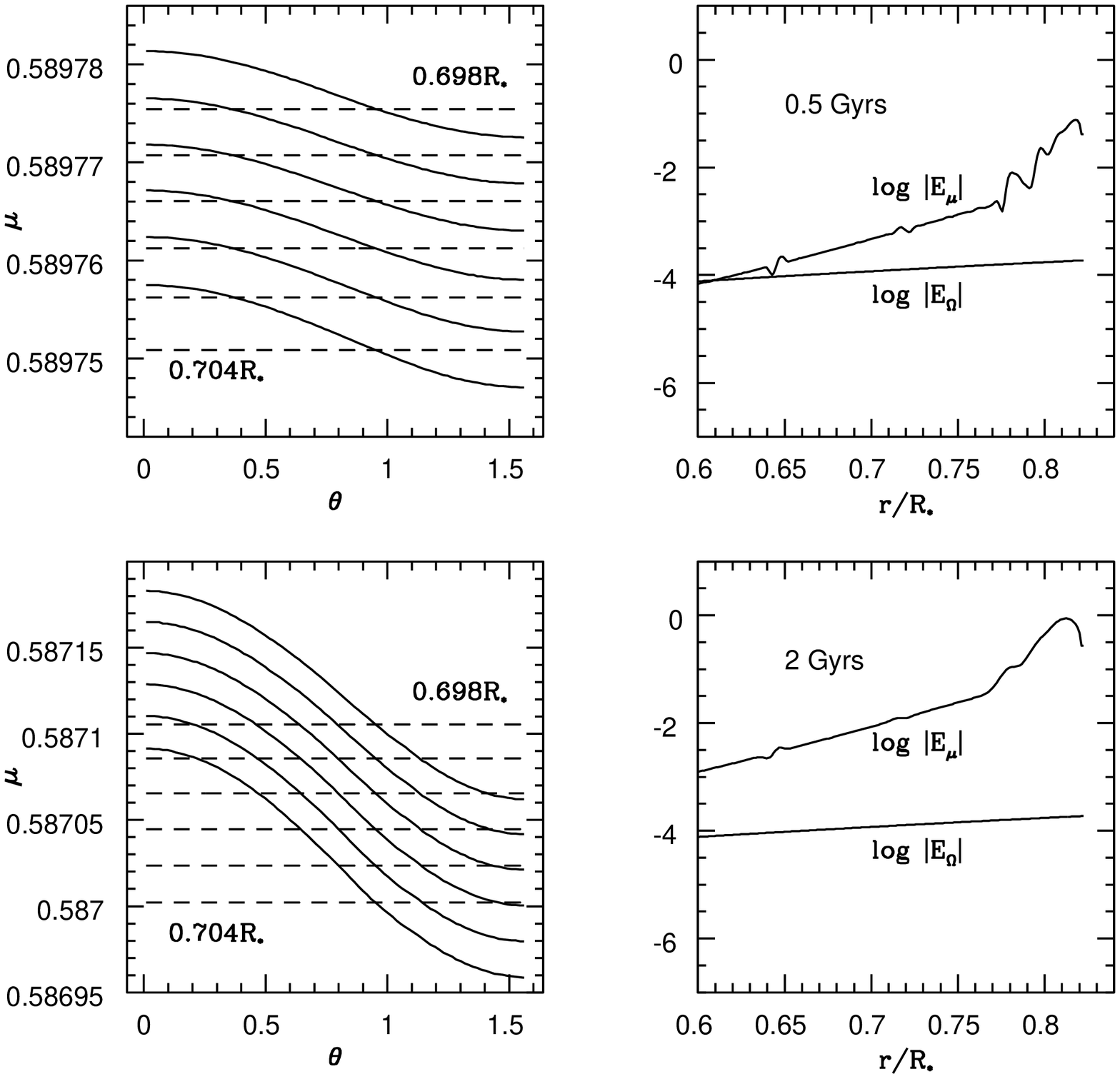}
\caption{Computations including the helium microscopic diffusion and the
classical meridional circulation : both the $E_{\Omega}$ and the
$E_{\mu}$ term are computed but only the $E_{\Omega}$ term is taken into
account in the meridional circulation velocity computations (equation
\ref{Ur})). Left figure : molecular weight versus colatitude ($\theta$
in radians) at different depth (between 0.704R$_{*}$ and 0.698R$_{*}$)
below the convective zone. Right : $E_{\Omega}$ and $E_{\mu}$ profiles
inside the star. The $E_{\mu}$ term, which is not introduced in the
advection computations, is evaluated by computing the horizontal
$\mu$-gradients present in the star and using the equation \ref{emu}.
After 0.5 Gyrs, the $E_{\mu}$ term is greater than the $E_{\Omega}$
term, therefore it cannot be neglected in the meridional velocity
computations.}
\label{fig5}
\end{figure}

\subsection{Importance of $\mu$-currents}
In the previous section we have presented the molecular weight
variations under the combined effects of helium diffusion and
$\Omega$-currents only. We have shown that diffusion builds vertical
$\mu$-gradients which are turned into horizontal ones by
$\Omega$-currents. Although the $\mu$-currents are not taken into
account in the computations of meridional advection, we have computed at
each time step the value of the $E_{\mu}$ term by determining the
horizontal $\mu$-gradients built in the star and using equation
\ref{emu}. We insist on the fact that these $\mu$-gradients are those
built by the classical meridional circulation, i.e. when the feed-back
effect of $\mu$-currents is not taken into account. 

In these conditions, figure \ref{fig5} (left column) displays the molecular
weight versus the colatitude at different depths below the convective
zone and at different steps of the computations. This shows the slow
construction of the horizontal $\mu$-currents. 
Figure \ref{fig5} (right column) compares the
$E_{\Omega}$ and $E_{\mu}$ terms inside the stars. It shows that the
horizontal $\mu$-gradients induced by the $\Omega$-currents lead rapidly
to important values of $E_{\mu}$. After 0.5 Gyrs, $E_{\mu}$ is already
greater than $E_{\Omega}$. The described situation is then unrealistic :
if such horizontal $\mu$-gradients were effectively built in the star,
the $\mu$-currents should have changed the sense of the circulation.
This clearly shows that the feed-back effect of $\mu$-currents cannot be
neglected in the computations.

\section{Meridional circulation including the $\mu$-currents}
\subsection{The creeping paralysis}
We now present the results obtained by introducing in the simulation the
diffusion of helium and the two currents of meridional circulation
($\Omega$ and $\mu$-currents). The advection by the circulation is then
computed by including in the computations of the meridional velocity the
classical term, $E_{\Omega}$ but also the $E_{\mu}$ term related to the
molecular weight gradients. 

\begin{figure}[h]
\epsscale{0.6}
\plotone{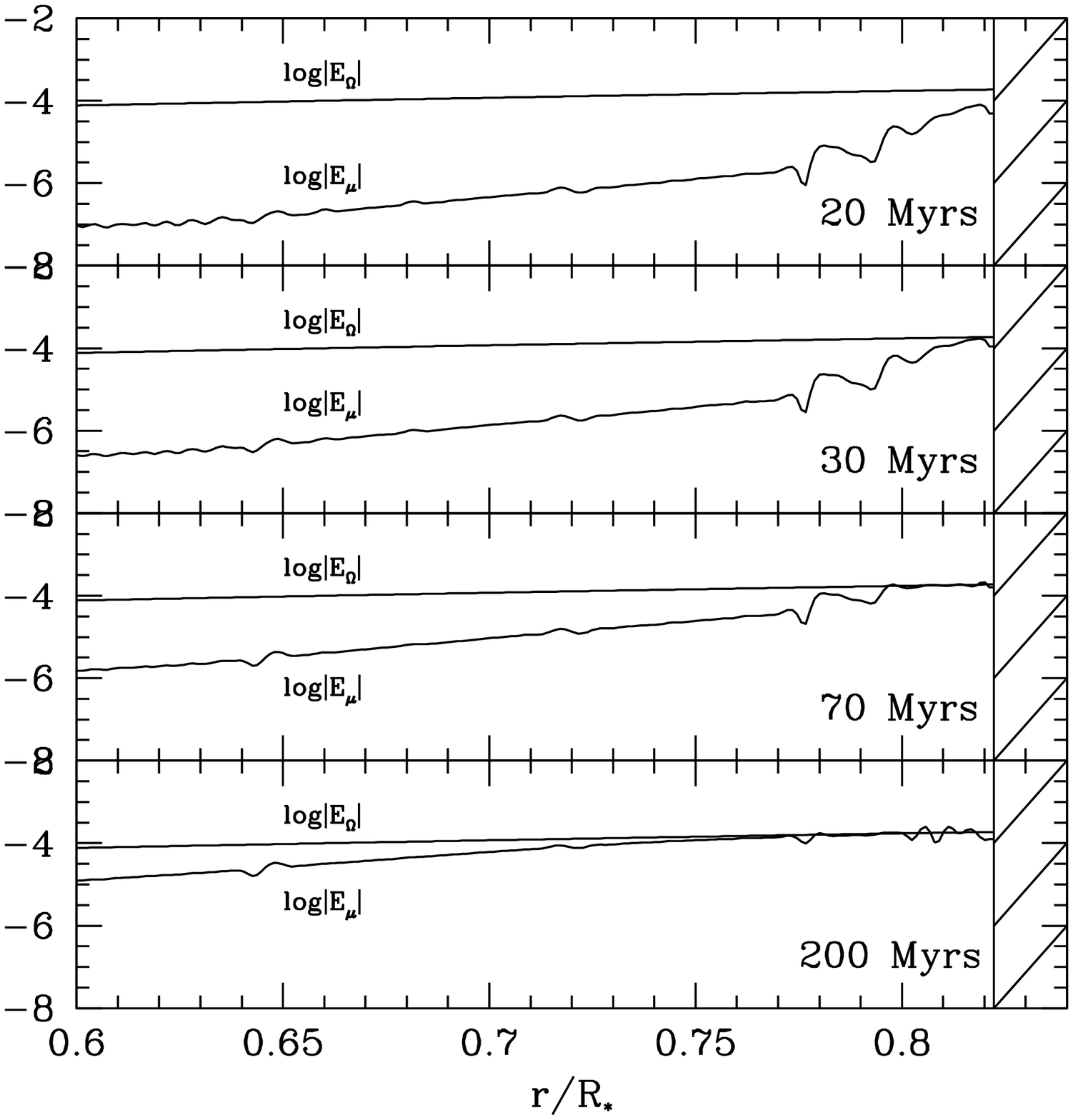}
\caption{Computations including He diffusion, $\Omega$-currents and
$\mu$-currents : $E_{\Omega}$ and $E_{\mu}$ profiles below the
convective zone at different steps of the simulation. Under the combined
effects of microscopic diffusion and meridional circulation, the
$\mu$-currents increase progressively in the star. They become rapidly
of the same order of magnitude as $\Omega$-currents below the convective
zone, which freezes out the circulation in the concerned region. The
paralyzed zone deepens then slowly inside the star.}
\label{eomega41}
\end{figure}

Figure \ref{eomega41} displays the $E_{\Omega}$ and $E_{\mu}$ profiles
inside the stars at different steps of the simulation. The
$\mu$-currents become rapidly of the same order of magnitude as the
$\Omega$-currents below the convective zone. As soon as the two currents
compensate each other, the meridional circulation seems to freeze out in
the concerned regions. The frozen region first appears below the
convective zone and deepens slowly inside the radiative zone. This is a
``creeping paralysis" process similar to that discussed by Mestel and
Moss \cite{Mestel86} for nuclear-induced $\mu$-gradients.

However the paralysis cannot be complete because of helium settling.
When the horizontal $\mu$-gradient, $\Lambda$, is just equal to the
critical value, $\Lambda_{crit}$, for which $|E_{\mu}|$ is equal to
$|E_{\Omega}|$, the meridional circulation vanishes. However helium
diffuses out of the convection zone where it is completely homogeneous
(i.e.  $\Lambda=0$). Because of this diffusion, the horizontal
$\mu$-gradients decreases below the convective zone and the circulation
is triggered again until a new equilibrium state is reached. The
$E_{\Omega}$ and $E_{\mu}$ profiles, which always remain very close in
the frozen region, suggest that the balance restoring time scale below
the convective zone is much more rapid than the diffusion time scale.
This is the reason why this problem is difficult to treat numerically.

\subsection{Self-regulating process}
The aim of this section is to understand more accurately what happens
when the balance between $E_{\Omega}$ and $E_{\mu}$ is broken due to the
helium microscopic diffusion below the convective zone. Then, for
numerical reasons, it is necessary to separate the computations of
diffusion and circulation in the simulation because the balance
restoring time scale is small compared to the diffusion time scale. In
other words, we need to let diffusion proceed long enough to create a
real unbalance, well above the numerical fluctuations, before studying
the restoring effect of the circulation.

We start this study from a situation where $|E_{\Omega}|=|E_{\mu}|$
below the convective zone in a significant region. We then stop the
meridional circulation and let helium diffusion proceed further until
the equilibrium between the two currents is clearly broken below the
convective zone. Afterwards we stop the diffusion and introduce the
circulation again in order to study its behavior.

\subsubsection{Computations}
\begin{figure}[h]
\epsscale{0.8}
\plotone{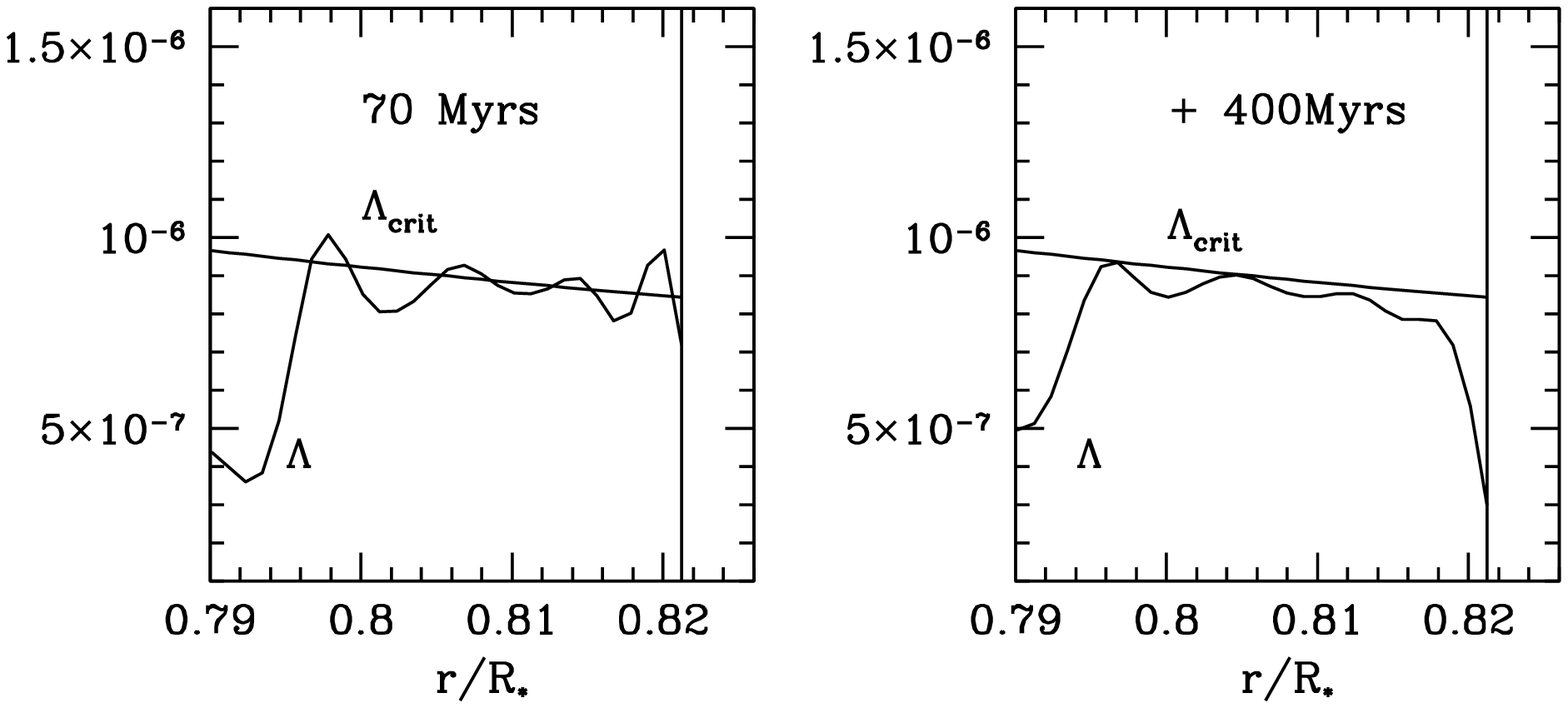}
\plotone{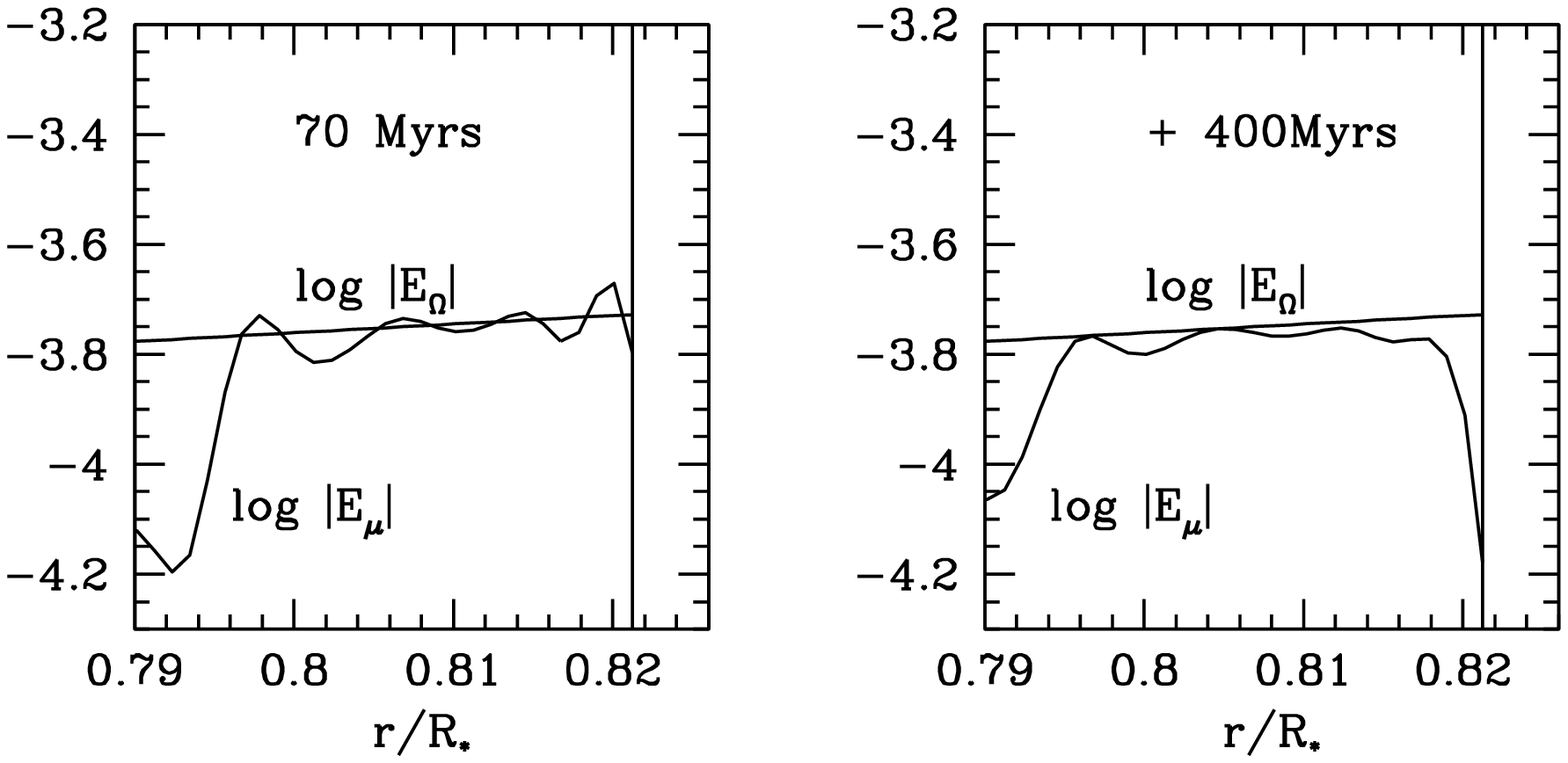}
\plotone{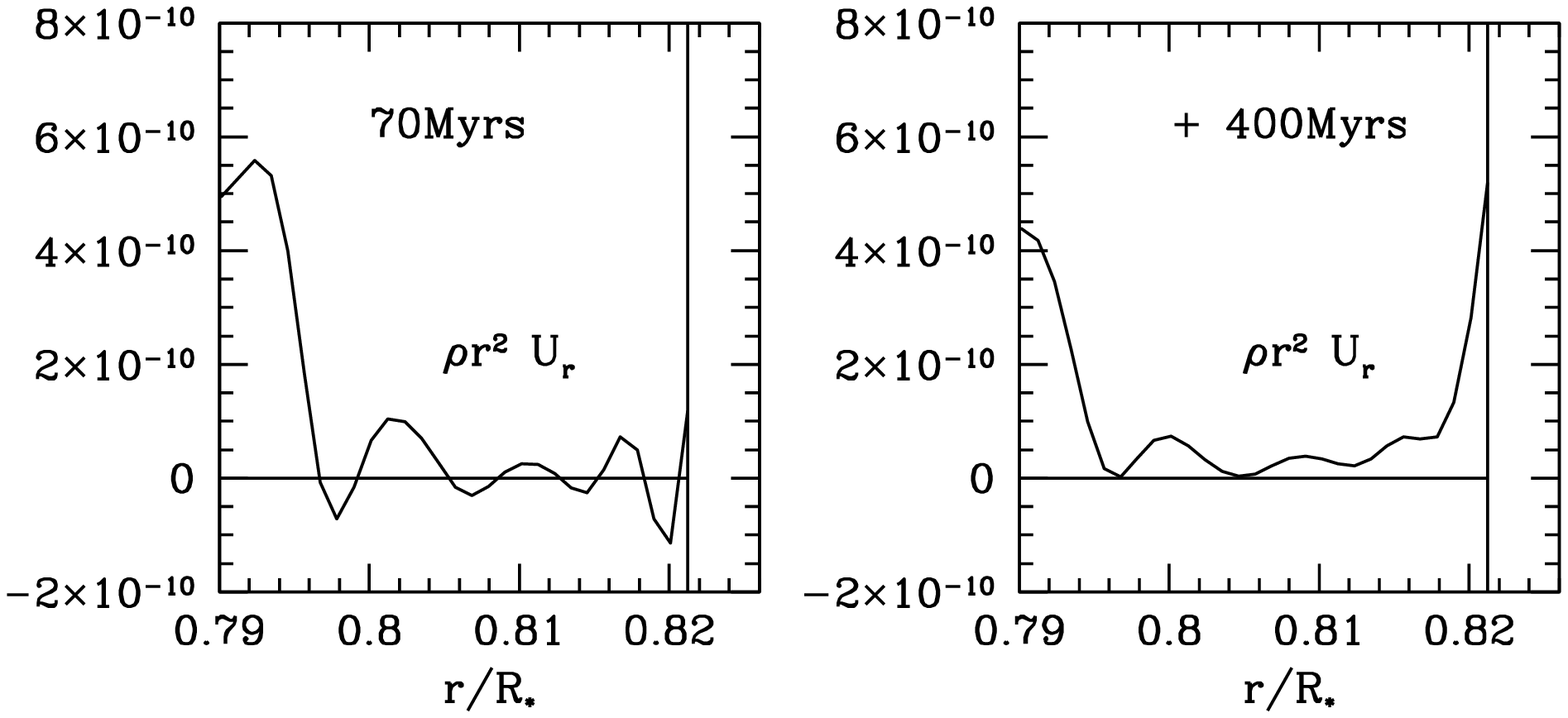}
\caption{Influence of microscopic diffusion on the ``creeping paralysis
" below the convective zone. Left : the figures show the results
obtained with the simulation (including diffusion + $E_{\Omega}$ +
$E_{\mu}$ ) after 70 Myrs. Then we stop the circulation and let the
diffusion proceed further during 400 Myrs. After this time, we stop the
diffusion and introduce the circulation again. Right : the figures
display the meridional circulation variables after the alteration of the
equilibrium by diffusion.}
\label{l1}
\end{figure}

As an initial situation, we use the results obtained with the complete
simulation after 70 Myrs (including diffusion and circulation, $\Omega$
and $\mu$-currents). This situation is represented by the third graph on
Figure 6 : the circulation is ``paralyzed" below the convective zone
down to a radius about $r/R_* = 0.795$. Then we stop the circulation
currents, as if they were completely frozen, and we let diffusion
proceed further by itself. Due to helium gravitational settling,
horizontaly homogeneous matter falls from the convective zone into the
radiative zone. We present here the results obtained if we let the diffusion proceed alone during 400 Myrs. The upper panel of figure \ref{l1} shows the $\Lambda$
profiles below the convective zone after the 400 Myrs of pure diffusion. Due to diffusion-induced homogeneization, a
significant decrease of the horizontal $\mu$-gradient may be observed
below the convective zone ($\Lambda < \Lambda_{crit}$) while $\Lambda$
remains close to $\Lambda_{crit}$ in the deeper layers down to the
boundary of the ``paralyzed" region. 

We then compute again the circulation currents. The middle panel of
figure \ref{l1} displays the $E_{\Omega}$ and $E_{\mu}$ profiles
obtained at that time. 
As expected, the balance between the two currents is clearly broken
below the convective zone : the fact that the horizontal $\mu$-gradient
has been forced below the critical value induces that $|E_{\mu}| <
|E_{\Omega}|$.

As a consequence, meridional circulation should settle again, with a
radial velocity amplitude given by the usual $U_r$ expression (equation
8). 

The lower panel of figure \ref{l1} displays, as a function of the
fractional radius, the radial flux of matter ($\rho r^2 U_r$) induced by
this circulation. According to equation \ref{ut}, the slope of this
profile determines the sign of the horizontal component of the
meridional circulation. In the radiative interior ($\rho r^2 U_r$)
decreases with radius, which means that the horizontal velocity is
positive, i.e. directed from the upward flow towards the downward flow
as usual. On the other hand, just below the convective zone ($\rho r^2
U_r$) increases up. Therefore, the horizontal circulation velocity is
negative in this region, going from the downward flow towards the upward
flow. 

\begin{figure}[h]
\plotone{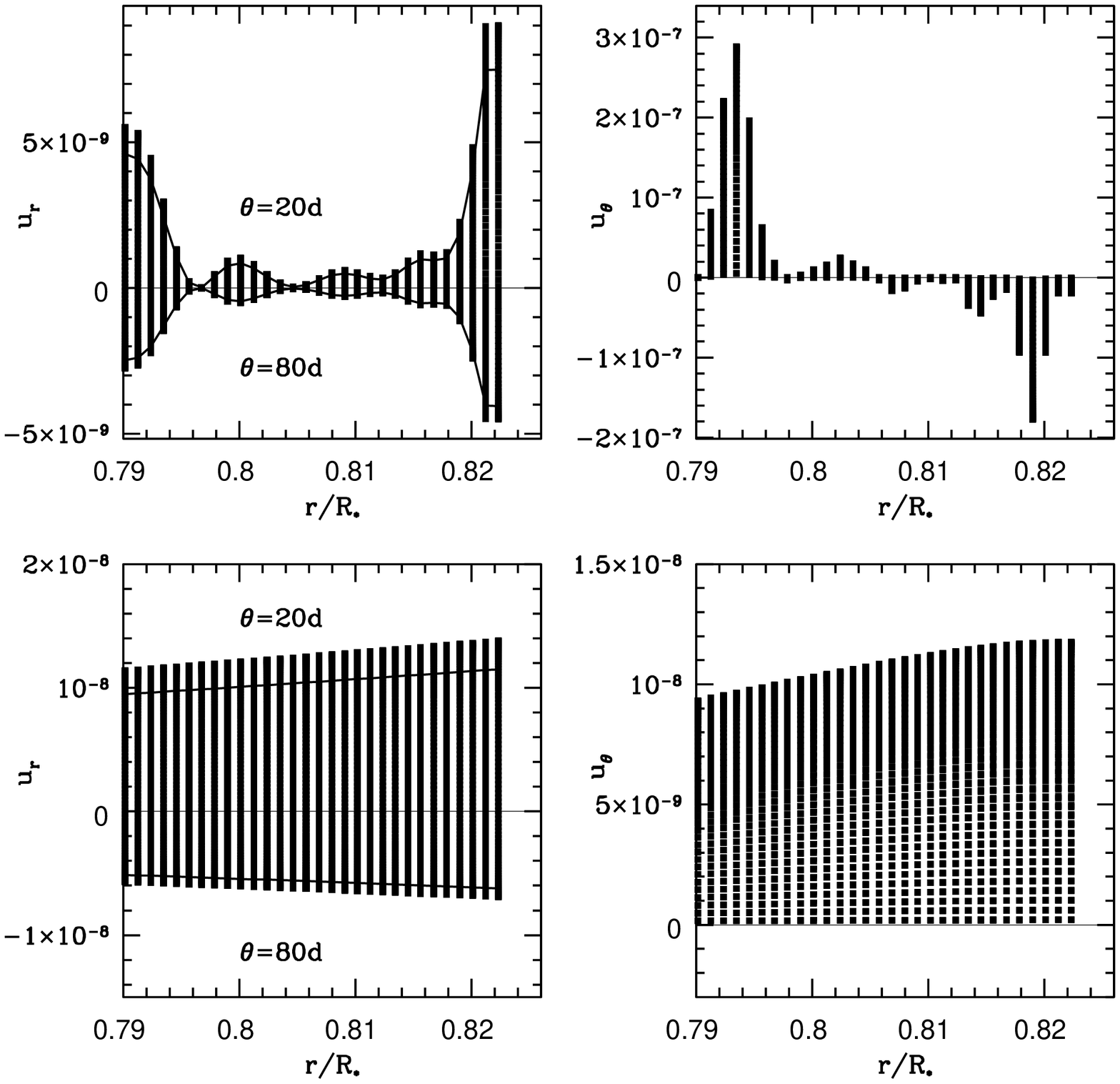}
\caption{Meridional velocity components $u_r$ (left) and $u_{\theta}$
(right) below the convective zone after 400 Myrs of pure microscopic
diffusion, compared to those obtained in the case of classical
meridional circulation. Each cell of the meshpoint is represented by a
black square, so that the velocities are shown as a function of the
fractional radius, for all angles $\theta$.
The upper and lower graphs are dramatically different. The upper graphs
clearly show the effect of the ``creeping paralysis" perturbated by
microscopic diffusion just below the convective zone.}
\label{vitesse}
\end{figure}

Figure \ref{vitesse} (upper panel) displays $u_r$ and $u_{\theta}$ below
the convective zone for each cell of the meshpoint in the present
situation and is compared to the case of the classical meridional
circulation (lower panel). The graphs show the radial velocity $u_r$
(left) or the latitudinal velocity $u_{\theta}$ (right)
as a function of the fractional radius, for all angles $\theta$.

Several interesting features arise. While the velocities are completely
smooth in the classical case, three regions appear for both components
in the present situation. In the deeper layers, the radial velocity is
positive in a sectorial part of the star (the upward flow) and negative
in the other part (downward flow) while the horizontal velocity is
always positive. Above these layers the two velocities become very small 
(``frozen region"). Finally, below the convective zone, the radial velocity
becomes again positive in one part and negative in the other part, but
the horizontal velocity is negative. Concerning the orders of magnitude,
in the present situation the vertical velocity is somewhat reduced
compared to the classical case, even at its maximum, while the magnitude
of the horizontal velocity is significantly enhanced above and below the
``frozen region".

\begin{figure*}
\epsscale{1}
\plottwo{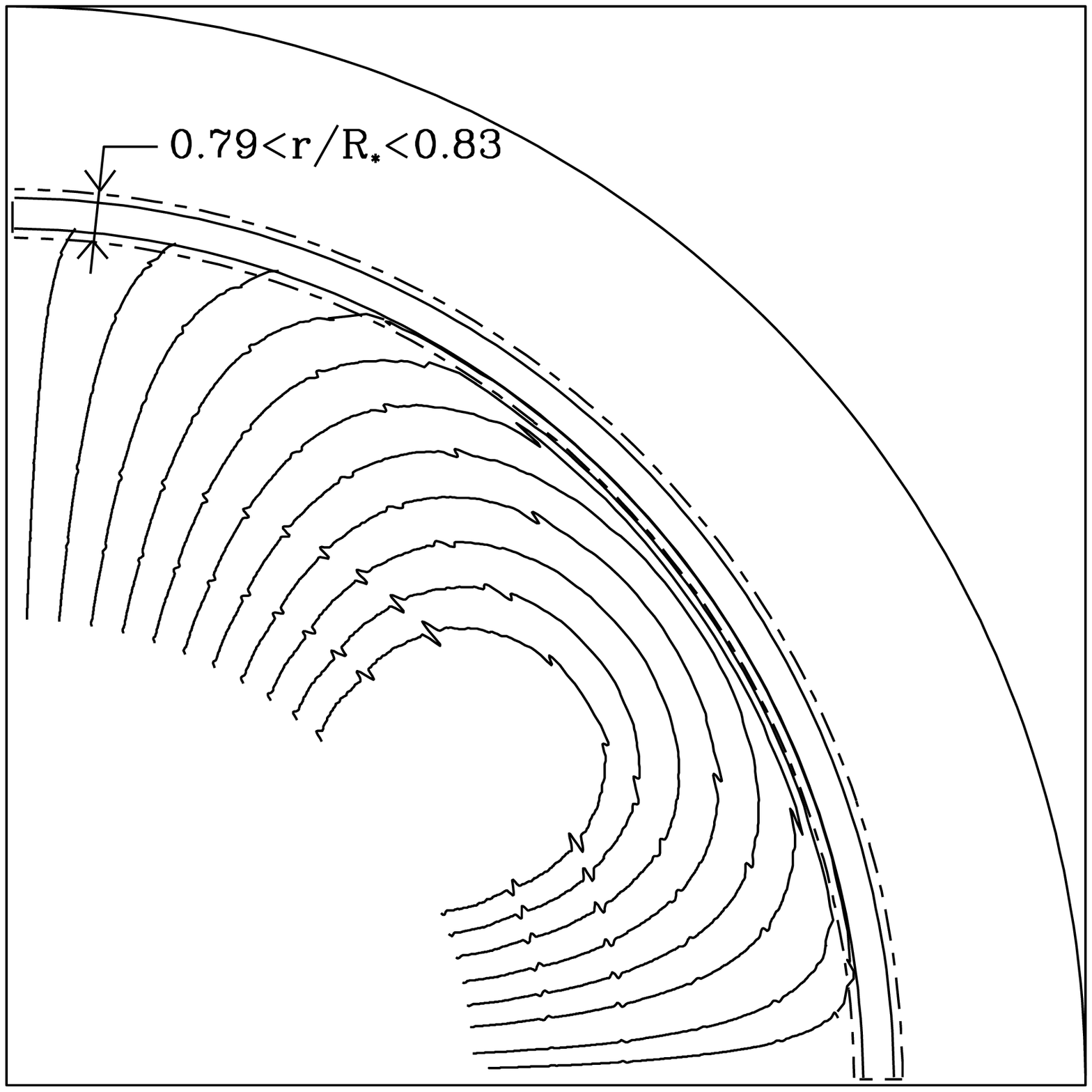}{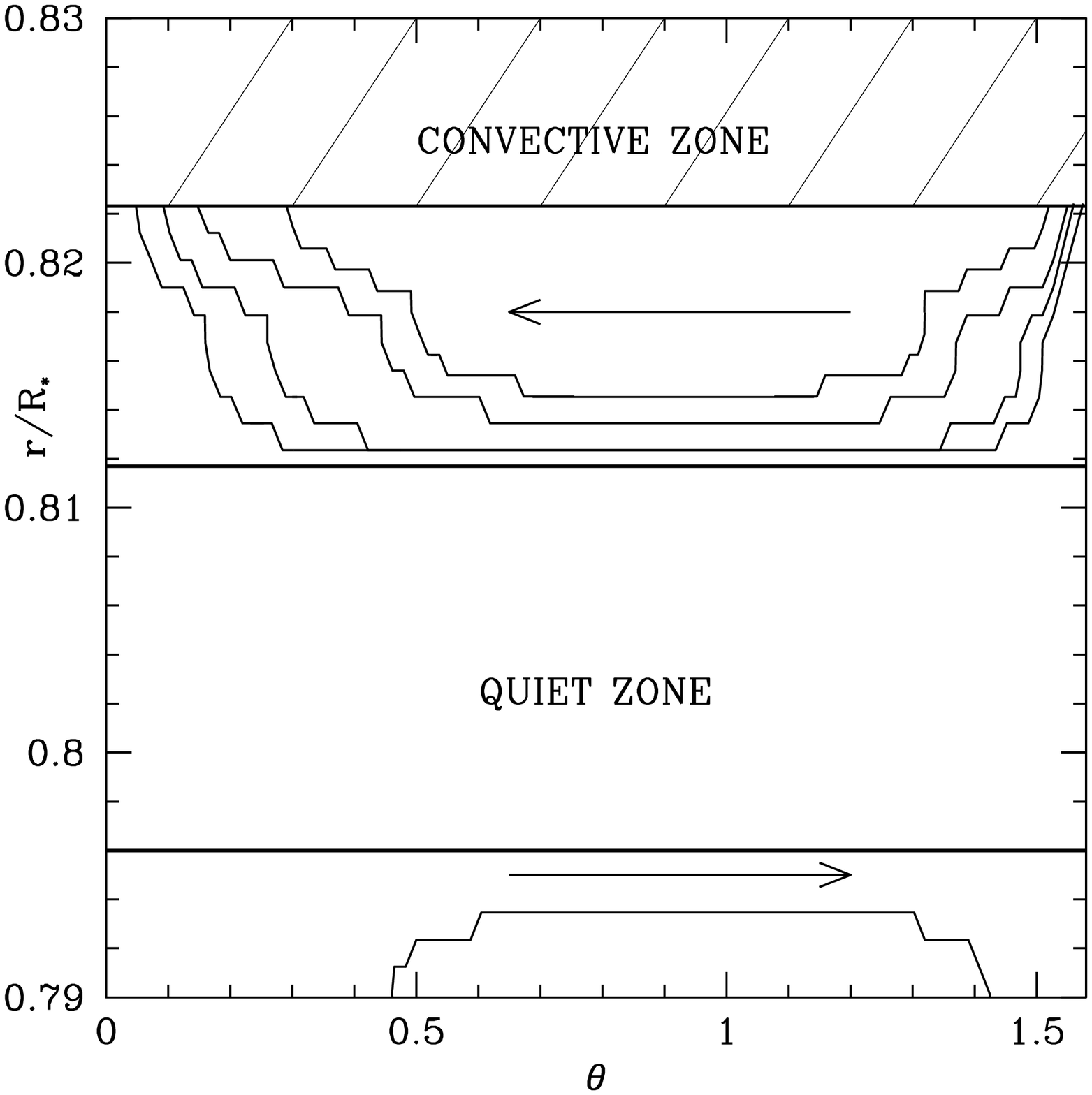}
\caption{The self-regulating process. During the simulation, under the
combined effects of helium diffusion and meridional circulation, the
$\mu$-currents increase. When they become equal to $\Omega$-currents,
the circulation vanishes. This phenomenon first occurs just below the
convective zone and a frozen region appears. The left figure shows the
meridional circulation streamlines obtained with the simulation after 70
Myrs. In the radiative interior, the circulation goes on in the
classical direction : from the pole to the equator. Below the convective
zone, the two meridional currents compensate each other and a ``frozen"
region has settled down in the star. Then diffusion proceeds further and
homogeneous matter falls from the convective zone into the ``frozen
region". It decreases the $\mu$-gradients and breaks up the balance
between the meridional currents. The figure on the right is a zoom of
the frozen region (the abscissa is $\theta$ in radians). It shows the
behavior of meridional circulation when the equilibrium between the two
currents has been broken by microscopic diffusion during 400 Myrs. Two
circulation loops appear. In the deep interior the meridional
circulation proceeds normally, but just below the convective zone a
second circulation loops settles from the equator to the pole. The two
circulation loops are separated by a region where the circulation is
very weak.}
\label{boucle}
\end{figure*}

Figure \ref{boucle} presents the resulting circulation streamlines in
the deep interior and just below the convective zone. The left panel
shows the circulation lines in a meridional sector after 70 Myrs of
complete simulation (diffusion and circulation). The ``frozen region"
clearly appears below the convective zone. In the radiative interior,
the circulation goes on in the classical direction but is compressed and
annihilated at the bottom of the ``frozen region". In the right panel
the region just below the convective zone is zoomed and expanded in
$\theta$, after 400 Myrs of pure microscopic diffusion inside the
``frozen region" . As expected from Figure \ref{vitesse}, two
circulation loops arise. In the radiative interior, the circulation
remains as seen on the left panel while the second loop, which appears
just below the convective zone, is clearly seen. The two loops are
separated by a quiet zone with an extension $r/R_* \simeq 3\%$

\subsubsection{Discussion}
We have shown that, when the ``frozen" state is altered by helium
microscopic diffusion below the convective zone, the circulation starts
again in order to compensate the created imbalance. A new circulation
loop appears, opposite to the classical one. We find numerically a similar
situation as described analytically in paper I. In this simulation, the two
loops are separated by about half a pressure scale height. We must not forget however that this is not the real situation which occurs in stars as in the present case the circulation has been completely stopped during 400 Myrs. 

The restoring time scale is very short compared to the diffusion one. If
we let the circulation start again after 400 Myrs of pure helium
settling, the region polluated by diffusion just below the convective
zone is mixed up in less than 1 Myr due to the new loop. In consequence,
we may deduce that as soon as diffusion begins to modify the equilibrium
between the $\Omega$ and the $\mu$-currents, the original situation is
immediately restored. This
corresponds to the situation described analytically in paper I, in which
we showed that a stationary stage could take place, where the horizontal
$\mu$-gradients should remain constant as well as the vertical ones, as
they are consistently linked.

In the presented computations, we let the circulation start again after 400 Myrs of pure helium settling. Others computations have been done with different time scales. In any case the restoring time scale is always very much smaller than the pure diffusion time, which is the main conclusion of this study.

\section{Conclusion}
After paper I, where we had discussed in an analytical approximate way
the importance of diffusion-induced $\mu$-gradients on the meridional
circulation in slowly rotating stars, we presented here a complete 2D
simulation of the considered processes. We used a stellar model obtained
from our stellar evolution code, that we discretized in the latitudinal
direction. Then we computed simultaneously the helium diffusion below
the convective zone and the advection due to meridional circulation. The
influence of helium variations on the stellar structure was neglected
(static model) but the feed-back effect of these variations on the
meridional circulation was thoughroughly introduced, in a complete way.

We confirmed that, in a somewhat short time
scale compared to the main sequence lifetime, the so-called
$\mu$-currents become of the same order of magnitude as the
$\Omega$-currents, thereby creating a ``frozen" region where the
circulation does not proceed anymore.

However, when this occurs the equilibrium between the two opposite
currents is permanently destabilized by the helium settling below the
convective zone. Helium falling from the mixed outer regions induces a
decrease of the horizontal $\mu$-gradients, which remain below the
critical values for which the circulation is completely stopped. As a
consequence, a new circulation loop develops, which mixes up into the
convective zone the region polluated by diffusion.

We derive two important consequences of this study :
\begin{itemize}
\item the rotation-induced mixing which occurs in stellar
radiative zones may be dramatically modified by $\mu$-gradients with the
occurrence of disconnected loops of circulation,
\item a new mixing process appears, directly induced
and modulated by the microscopic diffusion. 
\end{itemize}

The results concerning the rotation-induced mixing may have
important consequences for the light elements which are easily destroyed
by nuclear reactions ( Th\'eado and Vauclair 2002, paper III).
On the other hand, the influence of these processes on the abundance
variations induced by diffusion is still not completely settled.
A more sophisticated simulation would be helpful to derive it more precisely. However
the results of the present 2D simulation are very encouraging. The
diffusion-induced mixing which was analytically discussed in
paper I is confirmed. It acts as a restoring
system for the element abundance variations and increase in a
significant way their settling time scales.
We suggest, as in paper I, that diffusion and mixing react in such a way
as to keep both the horizontal and vertical $\mu$-gradients constant in
the ``frozen" region, while they proceed freely below.

\end{document}